\documentclass[%
 reprint,
 amsmath,amssymb,
 aps,
superscriptaddress,
]{revtex4-1}
\usepackage{graphicx}
\usepackage{dcolumn}
\usepackage{bm}
\usepackage{xcolor}
\usepackage{physics}
\usepackage{tikz}
\usepackage[normalem]{ulem}

\newtheorem{theorem}{Theorem}
\newtheorem{lemma}{Lemma}
\newtheorem{definition}{Definition}

\begin{document}

\preprint{APS/123-QED}

\title{The Thermodynamic Limit of Spin Systems on Random Graphs}

\author{Amy Searle}
\affiliation{%
 Department of Physics, University of Oxford, OX1 3PU
}%
\author{Joseph Tindall}
\affiliation{%
 Center for Computational Quantum Physics, Flatiron Institute, 162 5th Avenue, New York, NY 10010
}%
\email[]{jtindall@flatironinstitute.org}

\date{\today}

\newcommand{\jt}[1]{{\color{blue}{[JT: #1]}}}

\begin{abstract}
We utilise the graphon---a continuous mathematical object which represents the limit of convergent sequences of dense graphs---to formulate a general, continuous description of quantum spin systems in thermal equilibrium when the average co-ordination number grows extensively in the system size. Specifically, we derive a closed set of coupled non-linear Fredholm integral equations which govern the properties of the system. The graphon forms the kernel of these equations and their solution yields exact expressions for the macroscopic observables in the system in the thermodynamic limit. We analyse these equations for both quantum and classical spin systems, recovering known results and providing novel analytical solutions for a range of more complex cases. We supplement this with controlled, finite-size numerical calculations using Monte-Carlo and Tensor Network methods, showing their convergence towards our analytical results with increasing system size.

\end{abstract}

\maketitle


\textit{Introduction} -- 
The physical properties of interacting systems are strongly affected by the connectivity of their components. For instance, network topology plays a decisive role in the rate of disease spreading in infectious disease models \cite{DiseaseGeometry} whilst systematic studies have been undertaken into the affect of connectivity on the synchronisation of oscillators \cite{SynchronisationTopology1, SynchronisationTopology2, SynchonisationTopology3, SynchronisationTopology4}.
\par
In interacting spin systems, the same ideas hold true: frustration causes the manifestation of exotic phases of matter such as a spin liquids \cite{Takano1996} and small-world effects alter the underlying universality class of ordered-disordered phase transitions \cite{Paulo2003}. The difficulty of solving the many-body problem (especially in the quantum regime), however, means a more general characterisation of how network topology influences strongly correlated systems is unknown.
\par When disorder is absent and the average co-ordination number becomes large, interacting many-body systems fall into the mean-field universality class and become amenable to simpler, mathematical and computational approaches \cite{DMFT1, DMFT2, DMFT3, MFT1, MFT2}. Despite mostly being applied to translationally-invariant systems, the mean field approach is known to be valid for an infinite multitude of networks, whether homogeneous or heterogeneous \cite{HeterogeneousMFT}. Whilst it is only exact in the thermodynamic limit and when the average co-ordination number grows proportionally with the system size \cite{MFTValidity}, mean-field theory can provide meaningful physical predictions for low-dimensional systems \cite{Ramakrishnan1995, Pandit2012}.
Within the field of graph theory, network structures---whether heterogeneous or homogeneous---with an extensive co-ordination number are well-characterised. Their thermodynamic limit is succinctly described by the graphon \cite{Graphon1, Graphon2}, a continuous mathematical object which represents the limit of a sequence of adjacency matrices as the number of vertices tends to infinity and the average co-ordination number grows extensively. 

\par
Here we utilise the graphon in the study of interacting spin systems. This allows us to formally take the thermodynamic limit and derive an exact, continuous theory for the limit of sequences of discrete Hamiltonian on graphs of increasing size and average co-ordination number.
Specifically, we take a very general spin Hamiltonian defined over an arbitrary graph and, for sequences of dense graphs whose limit is known to converge to a given graphon, derive a coupled set of integral equations which exactly describe the equilibrium physics of the limit of the corresponding sequence of Hamiltonians. The graphon forms the kernel in these integral equations and the Physics of the system can be directly studied as a function of this object.
\par Taking several classical and quantum example models we demonstrate the utility of these integral equations: i) verifying previous results on all-to-all spin systems, ii) proving the existence of a finite-temperature phase transition in the classical Ising model for any graphon and iii) deriving analytical solutions for the equilibrium observables of spin models on novel non-trivial, heterogeneous networks. We reinforce our analytical solutions with large-scale, finite-size Monte Carlo and Tensor Network simulations. Whilst the spin systems we treat in this work are commonly studied due to their relevance as models of real-world magnetism, they also find application in many other branches of science, including the political, social and biological sciences \cite{IsingPoliticalModel1, IsingSocialScience1, IsingGenetics1}.

 \textit{Hamiltonian} --
Our starting point is $L$ qubits placed on the $L$ vertices of a graph $G_{L}$. The graph is specified by an $L \times L$ symmetric adjacency matrix $A_{G_{L}}$ with elements $A_{v,v'}$ which dictate the (weighted) connections between vertices (qubits) $v,v' \in [1...L], [1...L]$. The Hamiltonian reads
\begin{equation}
    H(G_{L}) = \frac{1}{L}\sum_{\substack{v,v'=1 \\ v > v'}}^{L}A_{v,v'}\left(\sum_{\alpha= x,y,z}J^{\alpha}\hat{\sigma}^{\alpha}_{v}\hat{\sigma}^{\alpha}_{v'}\right) + \sum_{\substack{v \in V \\ \alpha= x,y,z}}h^{\alpha}\hat{\sigma}^{\alpha}_{v'}.
\label{Eq:Hamiltonian}
\end{equation}
with $J^{\alpha}, h^{\alpha} \in \mathbb{R}$, and $\hat{\sigma}^{\alpha}_{v}$ the Pauli spin operator acting along the $\alpha$-spin-axis on vertex $v$.
Our focus is on graphs where $0 \leq A_{v,v'} \leq 1$ and $\sum_{v,v'}A_{v,v'} \propto L^{2}$. When the graph is unweighted  ($A_{v,v'} \in \{0, 1\}$) then we refer to the graph as `dense' because the the average co-ordination number diverges with system size. The factor of $\frac{1}{L}$ in $H(G_{L})$ is necessary to ensure a finite, non-trivial energy density. A variety of well-known models---including the Curie-Weiss \cite{ClassicalIsingErdosRenyi} and Lipkin Meshkov Glick \cite{EELMG, LMGOutofEqm} models---are contained within our Hamiltonian. The restriction $0 \leq A_{v,v'} \leq 1$ and the homogeneous nature of the field strengths, however, precludes our Hamiltonian from including disordered models such as spin-glass systems \cite{Young1986, Yao2018}.
\par \textit{Theory} -- In this work, by utilising tools from graph theory and mean-field theory, we formulate an explicit, exact, continuous description of this system in thermal equilibrium in the thermodynamic limit. 
In order to describe our continuum formalism we must first introduce the concept of a graphon. This can be done by taking the vertices $v = 1... L$ of a graph $G_{L}$ and performing the change of variables: $x = v/L \in [0,1]$.  We then define $W_{G_{L}}(x,y) : [0,1]^{2} \rightarrow [0,1]$, a real symmetric stepped function over the unit square such that for a given $(x,y) \in I_{v} \times I_{v'}$, where $I_{v} = [(v-1)/L, v/L]$ then $W_{G_{L}}(x,y) = A_{v,v'}$. Equipped with a well-defined metric for the similarity of two graphs, it can be shown that for certain sequences of graphs $(G_{L})_{L \in \mathbb{N}}$ the limit $\lim_{L \rightarrow \infty}W_{G_{L}}(x,y)$ converges to a well-defined symmetric function $W(x,y)$ known as the `graphon' \cite{Graphon1, Graphon2}. In the Supplemental Material (SM) we discuss these metrics in detail and provide theorems on the convergence of graph sequences under these metrics.
Importantly, it is also possible to move in the opposite direction and given a graphon $W(x,y)$ construct sequences of finite graphs whose limit is $W(x,y)$. These finite graphs can be constructed via one of two methods: `stochastic' or `weighted' sampling of $W(x,y)$ and we use $G^{S}_{L}$ and $G^{W}_{L}$ to refer to their respective realisation over the vertices $v = 1....L$. They can be constructed by defining the quantity
\begin{equation}
    P_{v, v'} = L^2 \int_{I_{v} \times I_{v'}} W(x,y) dx dy, \qquad I_{v} = [(v-1)/L, v/L].
    \label{Eq:GraphonProbabilities}
\end{equation}

The adjacency matrix of the unweighted graph $G^{S}_{L}$ is then defined by setting $A_{v,v'} = 1$ with probability $P_{v,v'}$ and $A_{v,v'} = 0$ otherwise. The adjacency matrix of the weighted graph $G^{W}_{L}$ is defined by setting $A_{v,v'} = P_{v,v'}$. A given sequence of such realisations is guaranteed to converge to the graphon $W(x,y)$ in the limit $L \rightarrow \infty$ \footnote{We point out that the Hamiltonian in Eq. (\ref{Eq:Hamiltonian}) is invariant under $A_{G_{L}} \rightarrow A_{G_{L}}A$, $J^{\alpha} \rightarrow J^{\alpha}/c$. This degree of freedom on the graphon and the $P_{v,v'}$ is trivial as it does not affect its functional form --- which is what governs the resulting equilibrium physics.}.
\par With the definition of the graphon in hand, the central result of this paper can be presented.
\begin{theorem}
    Let $(G_L)_{L \in \mathbb{N}} = (G_1, G_2, ...)$ be a sequence of finite-size graphs generated as stochastic or weighted realisations of the graphon $W(x,y)$. Then for a given inverse temperature $T = 1/\beta$ the macroscopic properties of the equilibrium states of the sequence of Hamiltonians $H(G_{L})_{L \in \mathbb{N}} = (H(G_1), H(G_2), ...)$ converges and are determined by the solution of the following coupled integral equations
    \begin{equation}
      \lambda^{\alpha}(x) = - J^{\alpha}\int_{0}^{1}\frac{W(x,y)\lambda^{\alpha}(y)\tanh(\beta \Lambda(y))}{\Lambda(y)} dy + h^{\alpha},
    \label{Eq:ContinuousMFEquationsMain}  
    \end{equation}
with $\alpha = x,y,z$, $\Lambda(x) = +\sqrt{(\lambda^{x}(x))^{2}+(\lambda^{y}(x))^{2}+(\lambda^{z}(x))^{2}}$ and the three functions $\lambda^{\alpha}(x)$, with $\alpha \in \{x,y,z\}$, each being continuous, real-valued and defined over the domain $[0,1]$. 
\label{Eq:Theorem1}
\end{theorem}
In order to prove this theorem and arrive at Eq. (\ref{Eq:ContinuousMFEquationsMain}) we state the following intermediate theorem
\begin{theorem}
    Let $f(H) = -\frac{1}{L \beta}{\rm ln}({\rm Tr}(\exp(-\beta H)))$ be the free energy density of a $d^{L} \times d^{L}$ many-body Hamiltonian, with $\beta \in \mathbb{R}_{\geq 0}$ and $d$ the dimension of the local Hilbert space. Let $G_{L}^{S}$ and $G_{L}^{W}$ be the stochastic and weighted realisations on $L$ vertices of a graphon $W(x,y)$ respectively. For an arbitrary set of real, finite values for the parameters $\{J^{x}, J^{y}, J^{z}, h^{x}, h^{y}, h^{z}\}$ the following is true
   \begin{equation}
        \vert f(H(G_{L}^{S})) - f(H(G_{L}^{W})) \vert = \mathcal{O}(L^{-1/2}),
    \end{equation}
   which vanishes in the limit $L \rightarrow \infty$.
   \label{Eq:Theorem2}
\end{theorem}
This theorem is a significant generalisation of theorem 1 in Ref. \cite{Tindall2022} which proved this result solely for sequences of Erdős-R\'enyi graphs, which correspond to the constant graphon. The proof of theorem \ref{Eq:Theorem2} (which can be found in the SM) relies on more general statistical properties of random graphs.  
\par With theorem \ref{Eq:Theorem2} in hand, theorem \ref{Eq:Theorem1} follows by: i) focusing strictly on the sequence $(G^{W}_L)_{L \in \mathbb{N}} = (G^{W}_1, G^{W}_2, ...)$ of weighted finite realisations of $W(x,y)$, ii) applying mean-field theory (which is exact here in the thermodynamic limit) and iii) taking the continuum limit of the resulting equations by invoking the definition of the graphon. The SM contains full proofs of both theorem \ref{Eq:Theorem1} and theorem \ref{Eq:Theorem2}.
\par If we can solve Eq. (\ref{Eq:ContinuousMFEquationsMain}) for the functions $\{\lambda^{x}(x), \lambda^{y}(x), \lambda^{z}(x)\}$, then we have determined the equilibrium physics of the limit of the sequence $(H(G_1), H(G_2), ...)$. The functions $\{\lambda^{x}(x), \lambda^{y}(x), \lambda^{z}(x)\}$ are a change of variables from the continuum limit of the spin degrees of freedom in the Hamiltonian. They directly encode the physical properties of the equilibrium state: the magnetisation on site $v$ in the thermodynamic limit is specified by $\langle \sigma^{\alpha}(x) \rangle$ with $x = \lim_{L \rightarrow \infty}\frac{v}{L}$ and is related to the $\lambda$ functions by
\begin{equation}
    \langle \sigma^{\alpha}(x) \rangle = -\frac{\lambda^{\alpha}(x){\rm tanh}(\beta \Lambda(x))}{\Lambda(x)}.
\end{equation}
The total magnetisation along a given spin direction is
\begin{equation}
    M^{\alpha} =\lim_{L \rightarrow \infty}\frac{1}{L}\sum_{v=1}^{L}\langle \sigma^{\alpha}_{v} \rangle =  - \int_{0}^{1}\frac{\lambda^{\alpha}(x){\rm tanh}(\Lambda(x))}{\Lambda(x)} dx.
    \label{Eq:MagnetisationCalculation}
\end{equation}
The validity of the mean-field approximation here means we can compute multi-point correlators as products of on-site expectation values.

\par How can we solve Eq. (\ref{Eq:ContinuousMFEquationsMain}) and find $\{\lambda^{x}(x), \lambda^{y}(x), \lambda^{z}(x)\}$? In general there is no analytical solution and we will be restricted to numerical methods. Nonetheless, there are certain cases where they can be solved analytically. Consider the case the graphon is degenerate, i.e. $W(x,y) = \sum_{i=1}^{n}f_{i}(x)f_{i}(y)$ where $n$ is finite and $f_{i}(x): [0,1] \rightarrow [0,1]$. Substitution into the above equation tells us $\lambda^{\alpha}(x) = \sum_{i=1}^{n}c^{\alpha}_{i}f_{i}(x)$ where $c^{\alpha}_{i}$ are real-valued coefficients which depend on the field strengths $h^{\alpha}$, couplings $J^{\alpha}$ and the inverse temperature $\beta$ but do not depend on $x$. These coefficients $c^{\alpha}_{i}$ are the solution of the set of $3n$ coupled equations which result from the substitution of $\lambda^{\alpha}(x) = \sum_{i=1}^{n}c^{\alpha}_{i}f_{i}(x)$ into Eq.(\ref{Eq:ContinuousMFEquationsMain}). For a given set of $J^\alpha$, $h^{\alpha}$ and value of $\beta$ we therefore have a closed form for $\lambda^{\alpha}(x)$ and various observables in the system. In our examples in the main text (further examples, including non-degenerate graphons are considered in the SM) we focus on $n = 1$ as they can be manipulated to yield closed forms for the equilibrium properties of the system.
\par \textit{Classical Ising model} --
We first set $J^{x} = J^{y} = h^{x} = h^{y} = h^{z} = 0$ and $J^{z} = -1$, realising the classical Ising model with zero field. Utilising ${\rm sgn}(z)\tanh(\beta |z|) = \tanh(z)$, our integral equations reduce to
\begin{equation}
    \lambda^{z}(x) = \int_{0}^{1}W(x,y)\tanh(\beta \lambda^{z}(y) ) dy.
    \label{Eq:ClassicalIsingEqn}
\end{equation}
The $\mathbb{Z}_{2}$ spin-flip symmetry is encoded in the fact that if $\lambda^{z}(x)$ is a solution to the equation then so is $-\lambda^{z}(x)$. Moreover, there is clearly always the trivial solution $\lambda^{z}(x) = 0 \ \forall x$ which corresponds to the disordered paramagnetic state with $0$ magnetisation. Applying Banach's fixed-point theorem \cite{BanachsFixedPointTheorem} to Eq. (\ref{Eq:ClassicalIsingEqn}) tells us that, with certainty, when $\beta < {\rm sup}_{x \in [0,1]}\int_{0}^{1}W(x,y)dy$ this is the only solution. For larger values of $\beta$, however, there exists a non-trivial solution which corresponds to a ferromagnetic phase. For instance, when $\beta \rightarrow \infty$ we have $\lambda^{z}(x) = \int_{0}^{1}W(x,y)dx \neq 0 \ \forall x$. Thus, following this analysis, we know that $\lambda^{z}(x, \beta)$ cannot be smooth and continuous over $x \in [0,1]$ and $\beta \in [0, \infty]$ and there must exist a finite-order transition between the ferromagnetic solution and the paramagnetic solution at some critical temperature. Our continuum description has therefore allowed us to prove the existence of a ferromagnetic--paramagnetic phase transition for the Ising model on \textit{any} dense graph---with a corresponding analytical upper bound on this temperature. A similar argument can be applied to a number of the limits of Eq. (\ref{Eq:Hamiltonian}).
\par Now let us treat some explicit examples. We first consider $W(x,y) = p$ whose stochastic realisations are $G_{\rm ER}(p)$: the Erdős-R\'enyi graph over $L$ vertices where each edge appears independently with probability $p$. Observe from Eq. (\ref{Eq:ClassicalIsingEqn}) that in this case $\lambda^{z}(x) = \lambda^{z} = p\tanh(\beta \lambda^{z})$ and is independent of $x$. Substituting this into Eq. (\ref{Eq:MagnetisationCalculation}) gives us the familiar self-consistent equation $M^{z} = \tanh(\beta p M^{z})$ for the magnetisation $M^{z}$ of the classical Ising model under the mean-field approximation. The edge probability $p$ re-scales the temperature in the all-to-all model and the randomness of the model has no effect on the macroscopic physics in the thermodynamic limit---a result which has been proven to be general for spin systems on Erdős-R\'enyi graphs \cite{ClassicalIsingErdosRenyi, Tindall2022}. 

\begin{figure}[!t]
    \centering
    \includegraphics[width =\columnwidth]{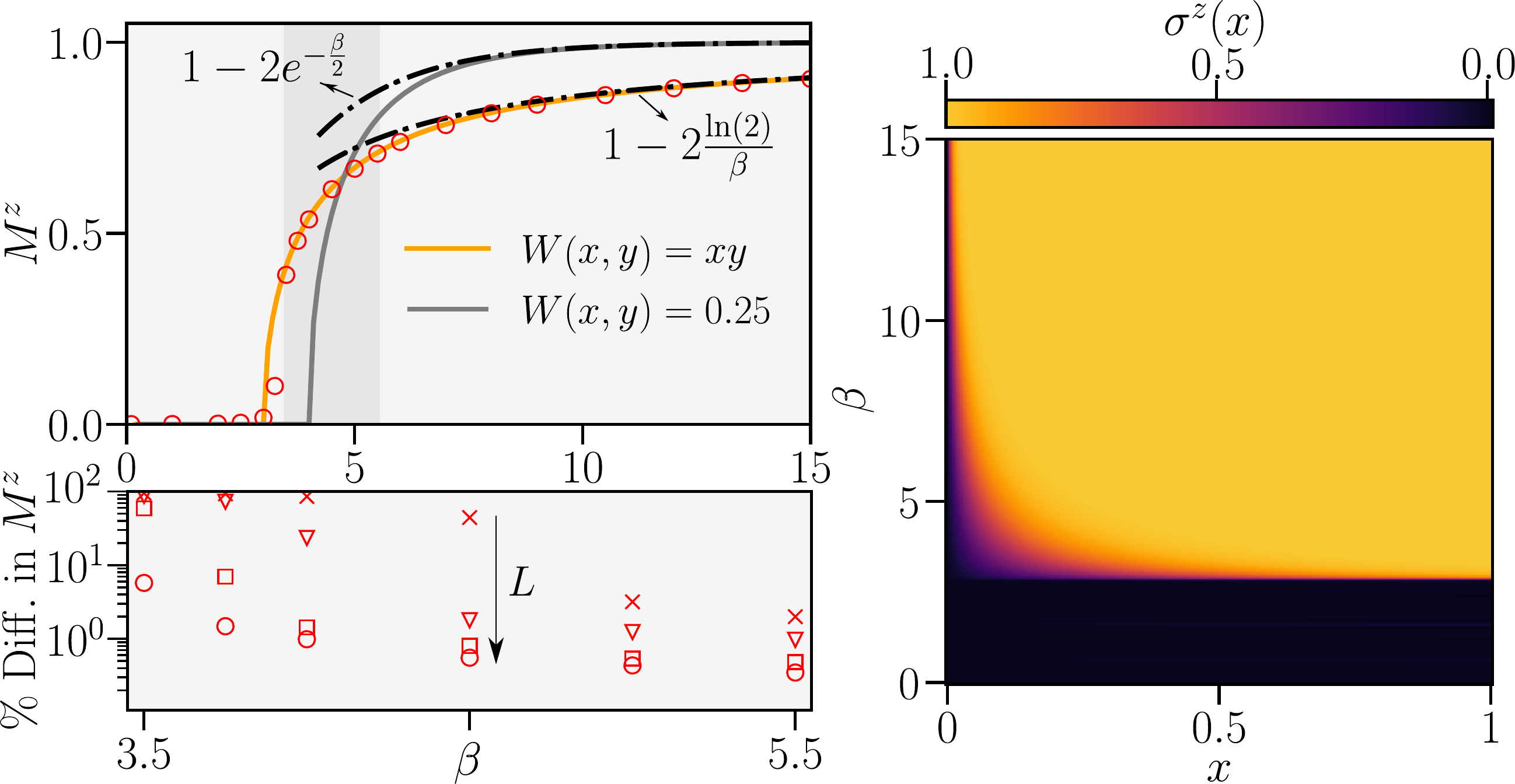}
    \caption{Magnetisation of the classical Ising model for the graphons: $W(x,y) = xy$ and $W(x,y) = \frac{1}{4}$. \textbf{a)} Total Magnetisation density $M^{z}$ versus inverse temperature $\beta$ for $L \rightarrow \infty$. Black dashed-dotted lines give the asymptotic derived by taking the large $\beta$ limit of the respective closed-form equations. Red circles correspond to Monte Carlo simulations of finite, $L=800$, randomly sampled graphs $G_{L}^{S}$ derived from $W(x,y) = xy$. Bottom) Percentage difference in $M^{z}$ for the exact result in the thermodynamic limit versus finite-size Monte Carlo simulations at several $L$ (crosses are $L = 100$, triangles $L = 200$, squares $L=400$ and circles $L = 800$). For each $\beta$ and $L$, $100$ stochastic samples $G^{S}_{L}$ are realised and the data (both top and bottom plots) is averaged over these. Further details are provided in the SM. \textbf{b)} On-site magnetisation $\sigma^{z}(x)$ versus $\beta$ and $x$ for the graphon $W(x,y) = xy$ in the thermodynamic limit.}
    \label{Fig:F1}
\end{figure}

\par We consider the, more complex, separable graphon $W(x,y) = xy$, whose stochastic relatisations dictates that each pair of spins $v$ and $v'$ interacts with a strength $1$ with probability $(vv'/L^{2})$ and strength $0$ otherwise. One can also choose to directly interpret the deterministic realisation of the graphon, where each pair of spins interacts with a strength $(vv'/L^{2})$. Both interpretations lead to the same physics in the thermodynamic limit---this follows directly from theorem \ref{Eq:Theorem2}. 
From Eq. (\ref{Eq:ClassicalIsingEqn}) we derive (see SM) $\sigma^{z}(x) = {\rm tanh}(\beta c x)$ and $M^{z} = \frac{{\rm ln}(\cosh(c))}{c}$ where $c$ is the  real-valued solution of the equation
\begin{equation}
    12c^{2}-\pi^{2}+24c {\rm ln}(1+e^{-2c}) - 12 {\rm PL}_{2}(-e^{-2c}) = \frac{24c^{3}}{\beta},
    \label{Eq:GraphonSolutionIsingXY}
\end{equation}
and ${\rm PL}_{2}$ is the PolyLogarithm function of order $2$. The critical inverse temperature $\beta_{c}$ is $\beta_{c} = 3$: the supremum of the LHS of the above equation for $c \in [0, \infty]$.

\begin{figure}[!t]
    \centering
    \includegraphics[width =\columnwidth]{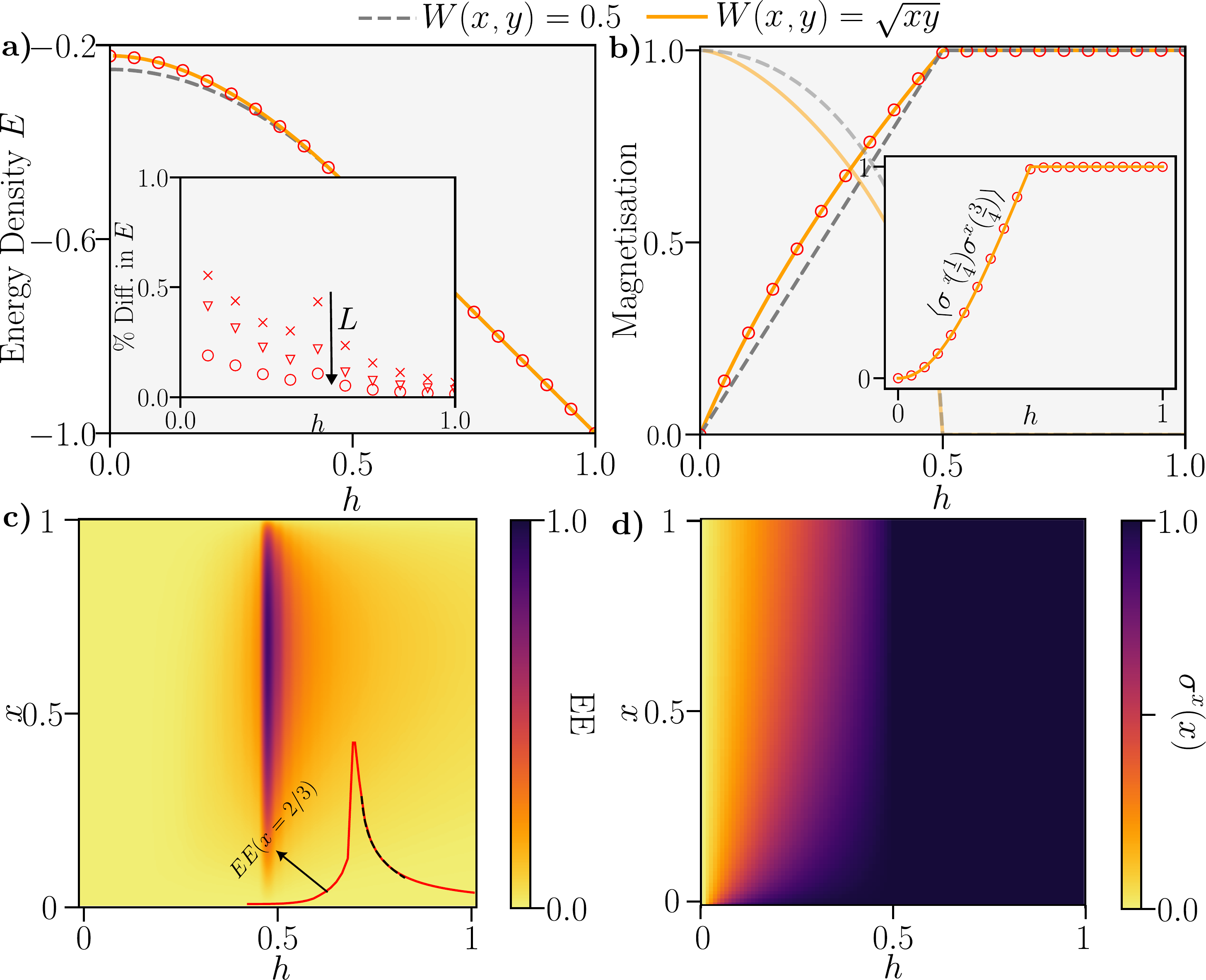}
    \caption{Properties of the ground state of the transverse field Ising model on the graphon $W(x,y)= \sqrt{xy}$. Results for the constant graphon are included for reference. \textbf{a)} Energy density versus transverse field strength $h$. Orange line represents the analytic solution in the thermodynamic limit. Markers represent numerical calculations averaged over $100$ finite stochastic realisations of $W(x,y) = \sqrt{xy}$ on $L = 400$ sites. Inset) Percentage difference between the ground state energy calculated on $L = 100, 200$ and $400$ (cross, triangle and circle marker respectively) site random-exchange realisations of $W(x,y)$ and the exact solution for $L \rightarrow \infty$. \textbf{b)} Total transverse (unfaded) and longitudinal (faded) Magnetisation densities of the ground state. Inset) Two-point correlator $\langle \sigma^{x}(\frac{1}{4})\sigma^{x}(\frac{3}{4}) \rangle$.
    \textbf{c)} Von-Neumann Entanglement Entropy (EE) of $W(x,y) = \sqrt{xy}$ averaged over $100$ stochastic realisations on $L = 400$ sites. The partition is between the first $xL$ sites of the system and the remaining $(1-x)L$ sites. The red curve corresponds to the entanglement entropy at $x = 2/3$ for $h = 0 \rightarrow 1$. Dotted black line is the fit ${\rm EE}(x = 2/3) = -0.136\log_{2}(h - 0.5) - 0.089$. \textbf{d)} Analytical result for the on-site magnetisation $\sigma^{x}(x)$ versus transverse field strength $h$ and position $x$ for the graphon $W(x,y) = \sqrt{xy}$.}
    \label{Fig:F2}
\end{figure}

\par In Fig. \ref{Fig:F1} we plot the total magnetisation $M^{z}$ and the local magnetisation $\sigma^{z}(x)$ versus $\beta$ based on our analytical solution. We also perform finite-size Monte-Carlo numerics for $M^{z}$ for increasing system size (by constructing stochastic realisations of $W(x,y)$) and demonstrate convergence to our analytical solution. We compare these results to the graphon $W(x,y) = \frac{1}{4}$. As the temperature increases both systems undergo a second-order phase transition characterised by typical mean-field exponents. For the graphon $W(x,y) = xy$, however, the convergence to a fully ferromagnetic state at zero temperature is slower. This convergence can be determined analytically by expanding Eq. (\ref{Eq:GraphonSolutionIsingXY}) for large $\beta$ and substituting into $M^{z} = \frac{{\rm ln}(\cosh(c))}{c}$, yielding $M^{z} = 1 - \frac{2\ln(2)}{\beta}$. There is thus a direct linear convergence of the magnetisation to unity with temperature $T = \frac{1}{\beta}$ versus the exponentially fast convergence associated with the homogoneous $W(x,y) = {\rm const.}$ case.
\par This slow convergence is a result of the `left boundary' of the system. In Fig. \ref{Fig:F1}b) we see that the local magnetisation at small values of $x$, where the spin-spin couplings are very weak, is very small even deep in the ferromagnetic regime. This `boundary effect' means the $T = 0$ state has a finite magnetic susceptibility to changes in temperature, i.e. $\frac{dM^{z}}{dT}\vert_{T=  0} = -2\ln(2)$. In the homogeneous case we have  $\frac{dM^{z}}{dT}\vert_{T=  0} = 0$. Whilst both systems are mean-field in terms of their universal behaviour, they exhibit very different physics in the ferromagnetic regime.

\par \textit{Transverse field Ising model} --
We now consider a quantum example: the transverse field Ising model. Our integral equation is (setting $J^{x} = J^{y} = h^{y} = h^{z} = 0$ and $h^{x} = -h, J^{z} = -1$ in Eq. (\ref{Eq:Hamiltonian})
\begin{equation}
    \lambda^{z}(x) = \int_{0}^{1}\frac{W(x,y)\lambda^{z}(y)\tanh(\beta \sqrt{h^{2} + (\lambda^{z}(y))^{2}})}{\sqrt{h^{2}+(\lambda^{z}(y))^{2}}} dy.
    \label{Eq:IsingEqn}
\end{equation}
We focus on the ground-state by taking the limit $\beta \rightarrow \infty$. We can again use Banach's fixed-point theorem here to prove the existence of a disordered-ordered phase transition with an upper bound of the critical field strength $h_{c}$ given by the supremum of the marginal of the graphon.
\par We now consider some specific examples. First, taking the Erdős-R\'enyi graphon $W(x,y) = p$ straightforwardly yields the solution $M^{z} = \sqrt{1-\frac{h^{2}}{p^{2}}}$ consistent with a rescaled TFI model with all-to-all coupling \cite{Tindall2022}.
\par There are, however, other, less trivial graphons for which an exact analytical solution for the ground state properties can be found. Consider the separable case $W(x,y) = \sqrt{xy}$. Some algebra on Eq. (\ref{Eq:IsingEqn}) (see SM) leads to $\langle \sigma^{x}(x) \rangle = \frac{h}{\sqrt{h^{2} + g^{2}x}} $ and $\langle \sigma^{z}(x) \rangle = -\frac{g \sqrt{x}}{\sqrt{h^{2} + g^{2}x}} $ with
\begin{equation}
    g = \begin{cases}
  \frac{\sqrt{2}}{3}\sqrt{1+(1-3h)\sqrt{1+6h}}  & h < \frac{1}{2}, \\
  \qquad \qquad \qquad 0 & \text{otherwise.}
  \label{Eq:RootXYg}
\end{cases}
\end{equation}
Integrating the expression (see SM) for the transverse magnetisation then gives the following closed form for the total transverse magnetisation density
\begin{equation}
    M^{x} = \begin{cases}
  \frac{6h}{3h + \sqrt{2 + 9h^{2} + (2-6h)\sqrt{1+6h}}}  & h < \frac{1}{2}, \\
  \qquad \qquad \quad 1 & \text{otherwise.}
  \label{Eq:RootXYTransverseMag}
\end{cases}
\end{equation}
The total longitudinal magnetisation density can also be obtained in closed form (see SM).
Our methodology has yielded an analytic expression for the magnetisation (in the thermodynamic limit) of the transverse field Ising model on a complex, highly inhomogeneous graph structure.

\par In Fig. \ref{Fig:F2} we plot these solutions alongside those for the constant graphon. The left boundary of the system, which has very weak $z-z$ coupling, modifies the physics of the system and makes it more susceptible to the transverse field than the all-to-all case. The transverse-field susceptibility versus site-index $x$ can be derived from Eq. (\ref{Eq:RootXYg}) yielding $\lim_{h \rightarrow 0}\frac{d \langle \sigma^{x}(x) \rangle}{dh}\vert = \delta(x)$, where $\delta(x)$ is the Dirac-delta function. There is a singularity in the susceptibility on the left boundary of the system at $0$ field-strength in the ferromagnetic regime. This is not present in the all-to-all model. Critical exponents for the magnetisations at the phase transition can be found via expansion of the analytical results and these are consistent with the mean-field universality class and equivalent for the two graphons. 
\par In Fig. \ref{Fig:F2} we also provide finite-size simulations of the ground state on random-exchange realisations of $W(x,y)$ using Density Matrix Renormalization Group (DMRG) \cite{DMRG} calculations on a Matrix Product State ansatz. We reach system sizes on the order of $\sim100$ spins, observing convergence to our analytical solution. We verify this convergence for local observables and non-local ones, where exact, analytical results can be obtained via the mean-field approximation. 
\par Importantly, from these tensor network numerics we can go beyond mean-field theory and obtain the entanglement entropy of the ground state on a finite system --- something currently inaccessible to our continuous formalism. This is non-zero and diverges logarithmically with the transverse field strength as criticality is approached: $h \rightarrow 0.5^{-}$. We also find the entanglement only depends on the ratio $x = N/L$, where $N$ is the partition size. This scaling is reminiscent of the entanglement properties of the all-to-all transverse field Ising model \cite{EELMG}. Here we observe it in a heterogeneous dense graph system, suggesting a possible universal mechanism underpinning the scaling of entanglement entropy in these models.

\par \textit{Conclusion} -- We have successfully utilised tools from graph theory to derive a set of integral equations which describe the physics of generic spin models with a large density of interactions in the thermodynamic limit--- whether classical or quantum. Our formalism straightforwardly reproduces known results and, most importantly, can be used to uncover the equilibrium properties of more complex systems. We observe how inhomogeneity in the underlying graphs alters the magnetic properties of the system.
\par Our work opens a up a number of further avenues for future research. Firstly, extending our formalism to describe the out-of-equilibrium dynamics of a spin system on a dense graph is a natural direction. Whilst an analytical solution is known for the all-to-all case ($W(x,y) = 1$) on the Lipkin-Meshkov-Glick model (a model whose dynamics was recently realised on a quantum simulator \cite{LMGOutofEqm}), our graph-theoretic approach could open up solutions for a whole range of dense graphs.  The quantum fluctuations which deviate finite-size results from the mean-field case would be stronger here.
\par Secondly, graphon estimation is the process of estimating the continuous graphon $W(x,y)$ from which a given finite graph $G$ could have been drawn from \cite{Stanley2014, Borgs2015, Jiaming2018}. Therefore when studying spin models on a large, connected structure (the structure need not necessarily be dense, graphon estimation can be done for quasi-sparse graphs too \cite{Borgs2015, Jiaming2018}) one can estimate the graphon $W(x,y)$ and solve our equations to obtain an approximate solution to the equilibrium physics of the system.

\par \textit{Acknowledgements} - JT is grateful for ongoing support through the Flatiron Institute, a division of the Simons Foundation. AS acknowledges support from
EPSRC Standard Research Studentship (Doctoral Training Partnership), EP/T517811/1, and the Smith-Westlake Graduate Scholarship at St. Hugh’s College. We are grateful to Sam Staton, Dieter Jaksch, Dries Sels and Vadim Oganesyan for fruitful discussions. Monte-Carlo calculations were performed with code written solely by the authors whilst Density-Matrix Renormalisation Group Calculations were done with the help of the ITensor library \cite{ITensor}.

\section{Appendix}

\renewcommand{\theequation}{S\arabic{equation}}
\renewcommand{\figurename}{Supplementary Figure}
\setcounter{equation}{0}
\setcounter{figure}{0}   
\setcounter{theorem}{0}   

\subsection{Appendix A: Proof of Theorems 1 and 2.}
We restate the Hamiltonian from the main text
\begin{equation}
    H(G_{L}) = \frac{1}{L}\sum_{\substack{v,v'=1 \\ v > v'}}^{L}A_{v,v'}\left(\sum_{\alpha}J^{\alpha}\hat{\sigma}^{\alpha}_{v}\hat{\sigma}^{\alpha}_{v'}\right) + \sum_{\substack{v \in V \\ \alpha}}h^{\alpha}\hat{\sigma}^{\alpha}_{v'},
\label{Eq:SMHamiltonian}
\end{equation}
where all definitions are retained and $\alpha = x,y,z$. We now prove Theorems $1$ and $2$ from the main text, which are restated below.

\begin{theorem}
    Let $(G_L)_{L \in \mathbb{N}} = (G_1, G_2, ...)$ be a sequence of finite-size graphs generated as stochastic or weighted realisations of the graphon $W(x,y)$. Then for a given inverse temperature $T = 1/\beta$ the macroscopic properties of the equilibrium states of the sequence of Hamiltonians $H(G_{L})_{L \in \mathbb{N}} = (H(G_1), H(G_2), ...)$ converges and are determined by the solution of the following coupled integral equations
    \begin{equation}
      \lambda^{\alpha}(x) = - J^{\alpha}\int_{0}^{1}\frac{W(x,y)\lambda^{\alpha}(y)\tanh(\beta \Lambda(y))}{\Lambda(y)} dy + h^{\alpha},
    \label{Eq:SMContinuousMFEquationsMain}  
    \end{equation}
with $\alpha = x,y,z$, $\Lambda(x) = +\sqrt{(\lambda^{x}(x))^{2}+(\lambda^{y}(x))^{2}+(\lambda^{z}(x))^{2}}$ and the three functions $\lambda^{\alpha}(x)$, with $\alpha \in \{x,y,z\}$, each being continuous, real-valued and defined over the domain $[0,1]$. 
\label{Eq:SMTheorem1}
\end{theorem}

\begin{theorem}
    Let $f(H) = -\frac{1}{L \beta}{\rm ln}({\rm Tr}(\exp(-\beta H)))$ be the free energy density of a $d^{L} \times d^{L}$ many-body Hamiltonian, with $\beta \in \mathbb{R}_{\geq 0}$ and $d$ the dimension of the local Hilbert space. Let $G_{L}^{S}$ and $G_{L}^{W}$ be the stochastic and weighted realisations on $L$ vertices of a graphon $W(x,y)$ respectively. For an arbitrary set of real, finite values for the parameters $\{J^{x}, J^{y}, J^{z}, h^{x}, h^{y}, h^{z}\}$ the following is true
   \begin{equation}
        \vert f(H(G_{L}^{S})) - f(H(G_{L}^{W})) \vert = \mathcal{O}(L^{-1/2}),
    \end{equation}
   which vanishes in the limit $L \rightarrow \infty$.
   \label{Eq:SMTheorem2}
\end{theorem}

We will first prove Theorem \ref{Eq:SMTheorem1} by assuming that Theorem \ref{Eq:SMTheorem2} is true. Then we will prove Theorem \ref{Eq:SMTheorem2} to complete the proof.

\par We first perform a mean-field treatment of $H(G_{L})$ for some arbitrary graph $G_{L}$ with adjacency matrix elements $A_{v,v'}$ and take $L \rightarrow \infty$. Let $\hat{\sigma}^{\alpha}_{v} = \langle \hat{\sigma}^{\alpha}_{v} \rangle  + \hat{\delta}^{\alpha}_{v}$, substitute it into the Hamiltonian and ignore terms of order $\hat{\delta}^{2}$. The result is (up to a constant):
\begin{equation}
    H(G_{L}) = \sum_{v}H_{v}= \sum_{v, \alpha}\hat{\sigma}^{\alpha}_{v}\Bigg(\frac{1}{L}\bigg(\sum_{v'=1}^{L}A_{v,v'}J^{\alpha}\langle \hat{\sigma}^{\alpha}_{v}\rangle \bigg) + h^{\alpha} \Bigg).
\end{equation}
Within this mean-field approximation the equilibrium state of the system is given by
\begin{equation}
    \rho(\beta) = \frac{\exp(-\beta H)}{{\rm Tr}(\exp(-\beta H))} = \bigotimes_{v=1}^{L}\frac{\exp(-\beta H_{v})}{{\rm Tr}(\exp(-\beta H_{v}))} = \bigotimes_{v=1}^{L}\rho_{v}.
\end{equation}
Where the reduced density matrix on each site $\rho_{v}$ is, explicitly (in the basis spanned by the eigenstates of $\sigma^{z}$), the following $2 \times 2$ matrix:
\begin{equation}
    \rho_{v} = \frac{1}{2\lambda_{v}}\begin{pmatrix}
        \lambda_{v}-\lambda_{v}^{z}{\rm tanh}(\beta \lambda_{v}) & -(\lambda_{v}^{x}-{\rm i}\lambda_{v}^{y}){\rm tanh}(\beta \lambda_{v}) \\
        -(\lambda_{v}^{x}+{\rm i}\lambda_{v}^{y}){\rm tanh}(\beta \lambda_{v}) & \lambda_{v} + \lambda_{v}^{z}{\rm tanh}(\beta \lambda_{v}),
    \end{pmatrix}
\end{equation}
where we have defined $\lambda_{v} = \sqrt{(\lambda^{x}_{v})^{2}+(\lambda^{y}_{v})^{2}+(\lambda^{z}_{v})^{2}}$ and $\lambda^{\alpha}_{v} = \frac{1}{L}\bigg(\sum_{v'}A_{v,v'}J^{\alpha}\langle \hat{\sigma}^{\alpha}_{v'}\rangle \bigg) + h^{\alpha}$.

\par By taking the expectation values $\langle \sigma^{\alpha}_{v} \rangle$ associated with $\rho_{v}$ we find the $\lambda^{\alpha}_{v}$ variables must obey the following self-consistency relation:
\begin{equation}
    \lambda^{\alpha}_{v} = -\frac{1}{L}\Bigg(\sum_{v'=1}^{L}\frac{J^{\alpha}A_{v,v'}\lambda^{\alpha}_{v}\tanh(\beta \lambda_{v'})}{\lambda_{v}}\Bigg) + h^{\alpha},
    \label{Eq:SMDiscreteMF}
\end{equation}
with $v = 1... L$ and $\alpha = x, y, z$.
The set of values $\{\lambda^{\alpha}_{v}\}$ with $v = 1...L$, $\alpha = x,y,z$ which solves the $3L$ non-linear equations described by Eq. (\ref{Eq:SMDiscreteMF}) thus fully characterise the mean-field equilibrium state associated with $H$.
\par Now we wish to take the continuum limit of Eq. (\ref{Eq:SMDiscreteMF}). First, we define the following: $x = v/L$, $dx = 1/L$, $\lambda^{\alpha}_{v} = \lambda^{\alpha}(x)$ and $\lambda_{v} = \lambda(x)$. We assume that the adjacency matrix has been generated as a weighted realisation of some graphon $W(x,y)$, i.e.  $A_{v,v'} =L^2 \int_{I_{v} \times I_{v'}} W(x,y) dx dy, \quad I_{v} = [(v-1)/L, v/L]$. Substituting this all into Eq. (\ref{Eq:SMDiscreteMF}) gives us
\begin{widetext}
\begin{equation}
            \lambda^{\alpha}(x) = -\Bigg(\sum_{y=1/L, 2/L, ..., L}\frac{J^{\alpha}L^2 \bigg( \int_{I_{v} \times I_{v'}} W(x,y) dx dy \bigg)\lambda^{\alpha}(x)\tanh(\beta \lambda(y))}{\lambda(y)}\Bigg)dx + h^{\alpha}.  
    \label{Eq:DiscreteMFV2}
\end{equation}
\end{widetext}
Now we take $L \rightarrow \infty$ which implies $L^{2}\bigg( \int_{I_{v} \times I_{v'}} W(x,y) dx dy \bigg) \rightarrow W(x,y)$ and the summation becomes an integral. We can then write down the coupled, continuous mean-field equations
\begin{equation}
      \lambda^{\alpha}(x) = - J^{\alpha}\int_{0}^{1}\frac{W(x,y)\lambda^{\alpha}(y)\tanh(\beta \Lambda(y))}{\Lambda(y)} dy + h^{\alpha},
    \label{Eq:ContinuousMFEquations}  
\end{equation}
with $\alpha = x, y, z$ and $\Lambda(x) = \sqrt{(\lambda^{x}(x))^{2}+(\lambda^{y}(x)^{2}+(\lambda^{z}(x))^{2}}$. These govern our system in the thermodynamic limit of the sequence of graphs generated from the graphon $W(x,y)$. Whilst we explcitly used the weighted realisation $G^{W}_{L}$ of $W(x,y)$, Theorem 2 tells us that the equilibrium properties of the system that arise as the solution of Eq. (\ref{Eq:ContinuousMFEquations}) are equivalent for both $G^{W}_{L}$ and $G^{S}_{L}$ as $L \rightarrow \infty$. Thus these equations govern the properties of \textit{any} sequence of finite graphs which converge to $W(x,y)$ --- not just weighted ones.
\par The equations in Eq. (\ref{Eq:ContinuousMFEquations}) are coupled, non-linear Fredholm integral equations with the graphon acting as the kernel. From the solution set $\{\lambda^{x}(x), \lambda^{x}(y), \lambda^{y}(z)\}$ to these equations we can obtain the on-site magnetisations via
\begin{equation}
    \sigma^{\alpha}(x) = - \frac{\lambda^{\alpha}(x)\tanh(\beta \Lambda(x))}{\Lambda(x)},
    \label{Eq:SMMagnetisationCalculation}
\end{equation}
and the total magnetisation density is $M^{\alpha} = \int_{0}^{1}\sigma(x)dx$.

In order to complete the proof of Theorem 1 we need to prove Theorem 2 which was assumed at the end of the last section.
We recall a Lemma proven in Ref. \cite{Tindall2022}, which we will be helpful in completing the proof. 
\begin{lemma}
Let $(A_{L})_{L \in \mathbb{N}} = (A_1, A_2,...)$ and $(B_{L})_{L \in \mathbb{N}} = (B_1, B_2,...)$ be two sequences of many-body Hermitian matrices. The matrices $A_L$ and $B_L$ in the sequence have size $d^L \times d^L$, with $d$ fixed and the dimension of the local Hilbert space. Let $D_L = A_L - B_L$ and $\lambda_{\rm Max}^D$ be the largest (in terms of the absolute value) eigenvalue of $D_L$. If $|\lambda_{Max}^D| = \mathcal{O}(L^{\kappa})$ then $|f(A_L) - f(B_L)| = \mathcal{O}(L^{\kappa - 1})$, which vanishes for $\kappa < 1$ as $L \rightarrow \infty$.
\end{lemma}

We begin by defining the following operator:
\begin{align}
    D_L &:= H(G_{L}^{S}) - H(G_{L}^{W}) = \sum_{\alpha = x,y,z} D_L^{\alpha} \notag \\ &= \sum_{\alpha = x,y,z}\frac{1}{L} \left( \sum_{\substack{i,j = 1 \\ i< j}}^{L} A^{S}_{ij} \hat{\sigma}_i^{\alpha} \hat{\sigma}_j^{\alpha}  -  \sum_{\substack{i,j = 1 \\ i< j}}^{L}  A^{W}_{ij} \hat{\sigma}_i^{\alpha} \hat{\sigma}_j^{\alpha} \right),
\end{align}
where $A^{S}_{ij}$ are the matrix elements of $G_{L}^{S}$ and $A^{W}_{ij}$ are the matrix elements of $G_{L}^{W}$, finite stochastic and weighted realisations of some graphon $W(x,y)$.

\par  We proceed to evaluate the eigenvalues of the operator $D^{\alpha}_L$. As such, consider its eigenstates $ \ket{\sigma_1^{\alpha}, ..., \sigma_L^{\alpha}}$, where $\hat{\sigma}^{\alpha}_i \ket{\sigma_1^{\alpha}, ..., \sigma_L^{\alpha}} = \mu_{i} \ket{\sigma_1^{\alpha}, ..., \sigma_L^{\alpha}}$, with $\mu_{i} = \pm 1$, depending on whether the $i$th spin is pointing `up' or `down' in that basis. We will define $\mu_{ij}:= \mu_i \mu_j$ and consider the eigenvalue
\begin{align}
\bra{\sigma_1^{\alpha}, ..., \sigma_L^{\alpha}} &D^{\alpha}_L  \ket{\sigma_1^{\alpha}, ..., \sigma_L^{\alpha}} = \notag \\ &\frac{1}{L} \left( \sum_{\substack{i,j = 1 \\ i< j}}^{L} A^{S}_{ij} \mu_{ij}  -  \sum_{\substack{i,j = 1 \\ i< j}}^L A^{W}_{ij} \mu_{ij}  \right),
    \label{Eq:SMDLEigenvalues}
\end{align}
We we will proceed to show that, independent of the eigenstate $ \ket{\sigma_1^{\alpha}, ..., \sigma_L^{\alpha}}$, this eigenvalue grows, at most, as $L^{1/2}$ in the large $L$ limit. From there we can invoke Weyl's inequality to show that the eigenvalues of $D_L$ grow asymptotically as $L^{1/2}$ and subsequently invoke Lemma $1$ (from above) to complete the proof of Theorem 1. 
\par To prove the $L^{1/2}$ growth of Eq. (\ref{Eq:SMDLEigenvalues}), we begin with the Hoeffding inequality. This states that for independent random variables $Y_1, ..., Y_n$ for which $a_i \leq Y_i \leq b_i$ then the sum $S_n:= Y_1+Y_2+...+Y_n$ is bounded as
\begin{equation}
    P(\left| S_n - \mathbb{E}(S_n) \right| \geq t) \leq 2 \exp(-\frac{2t^2}{\sum_{i=1}^n(a_i - b_i)^2}),
    \label{Eq:SMChernoff-Hoeffding}
\end{equation}
with the factor of 2 stemming from the fact we have incorporated both the upper and lower Hoeffding bounds together. We will apply this bound to Eq. (\ref{Eq:SMDLEigenvalues}).
\par Let us construct the set $X := \{X_{ij}\}$ which consists of the $\frac{L(L-1)}{2}$ random variables $X_{ij}:= A^{S}_{ij}\mu_{ij}$ (we have $i, j = 1... L$ and $i > j$).
Observe that since $A^{S}_{ij} \in \{0,1\}$ (the graph $G_{L}^{S}$ is simple) and $\mu_{ij}$ is either $1$ or $-1$, we have that $-1 \leq X_{ij}  \leq 1$. Also, $\mathbb{E}(X_{ij})= \mathbb{E}(A^{S}_{ij}) \mu_{ij} = A^{W}_{ij}\mu_{ij}$, where $\mathbb{E}(A^{S}_{ij})$ denotes the expected value. In our case, however, all of the $X_{ij}$ quantities are not independent--- the sign of $X_{ij}$ is determined from the sign of $X_{ik}$ and $X_{kj}$. This can be dealt with by applying the Hoeffding bound to the $X_{ij}$'s with positive and negative sign separately. 
\par A given eigenvector $\ket{\sigma^{\alpha}_1, ..., \sigma^{\alpha}_L}$ will consist of a number of spins pointing up ($\sigma_{i} = +1$) and and the remainder pointing down ($\sigma_{i} = -1$). Define $M = \sum_{k=1}^{L}\sigma_{k}$, then it can be checked that of the $\frac{L(L-1)}{2}$ parameters $\mu_{ij}$, $\frac{L^{2} + M^{2}-2L}{4}$ are positive and $\frac{L^{2} - M^{2}}{4}$ are negative. We therefore partition the set $X:= \{X_{ij}\}$ into two sets as follows
$$
\begin{cases}
x^{A}:= \{X_{ij} \vert \mu_{ij} = 1\},\\
x^{B}:= \{X_{ij} \vert \mu_{ij} = -1\}.\\
\end{cases}
$$
We would like also to keep track of the values $A^{W}_{ij}\mu_{ij}$, so we partition the set $\{\mu_{ij} A^{W}_{ij}\}$ into 
$$
\begin{cases}
\bar{x}^{A}:= \{\mu_{ij} A^{W}_{ij} \vert \mu_{ij} = 1\},\\
\bar{x}^{B}:= \{\mu_{ij} A^{W}_{ij} \vert \mu_{ij} = -1\}.\\
\end{cases}
$$
We can now invoke the Hoeffding bound on each set separately, since each set now consists on independent random variables. We have two sums on which to invoke the bound:
$S^{A}:= \sum_{l=1}^{\frac{M^{2}+L^{2}-2L}{4}}x^{A}_{l}$ and $S^{B}:= \sum_{l=1}^{\frac{L^{2} -M^{2}}{4}}x^{B}_{l}$, where we use $x^{A}_{l}$ and $x^{B}_{l}$ to refer to elements from $x^{A}$ and $x^{B}$ respectively. Likewise we will use $\bar{x}^A_i$ and $\bar{x}^B_i$ to refer to individual elements of $\bar{x}^{A}$ and $\bar{x}^{B}$ respectively. Then observe that, due to $\mathbb{E}(X_{ij})= A^{W}_{ij}\mu_{ij}$  we have $\mathbb{E}(S^{A}) = \sum_{l=1}^{\frac{M^{2}+L^{2}-2L}{4}}\bar{x}^{A}_{l}$ and $\mathbb{E}(S^{B}) = \sum_{l=1}^{\frac{L^{2}-M^{2}}{4}}\bar{x}^{B}_{l}$
\par Equation (\ref{Eq:SMChernoff-Hoeffding}) then gives two bounds:
$$
\begin{cases}
    P \left( \left\vert S^{A} - \mathbb{E}(S^{A}) \right\vert \geq t_1 \right) \leq 
 2\exp(\frac{-2t_1^2}{L^{2}+M^{2}-2L}),\\
    P \left( \left\vert S^{B} - \mathbb{E}(S^{B}) \right\vert \geq  t_2 \right) \leq 2\exp(\frac{-2t_2^2}{L^2-M^2}).
\end{cases}
$$
We combine these bounds to obtain:
\begin{align}
    P\left( \left \vert \sum_{\substack{i,j = 1 \\ i< j}}^{L}\mu_{ij}A^{S}_{ij} -\sum_{\substack{i,j = 1 \\ i< j}}^{L}\mu_{ij}A^{W}_{ij} \right \vert \geq t_{1} + t_{2} \right)  \notag \\ \leq 4 \exp(\frac{-2t_1^2}{L^{2}+M^{2}-2L}) \exp(\frac{-2t_2^2}{L^2-M^2}).
\end{align}
Since we fixed the magnetisation $M$ of the eigenstate in deriving the above bound, we should take the union bound over ${L \choose \frac{1}{2} (L+M)}$ eigenstates with magnetisation $M$. We will denote an eigenstate with magnetisation $M$  as  $\ket{\sigma_M}$, resulting in the new bound
\begin{widetext}
\begin{equation}
    \bigcup_{\vert \sigma_{M} \rangle}P\left( \left \vert \sum_{\substack{i,j = 1 \\ i< j}}^{L}\mu_{ij}A^{S}_{ij} -\sum_{\substack{i,j = 1 \\ i< j}}^{L}\mu_{ij}A^{W}_{ij} \right \vert \geq t_{1} + t_{2} \right) \leq 4 {L \choose \frac{1}{2}(L+M)}\exp(\frac{-2t_1^2}{L^{2}+M^{2}-2L}) \exp(\frac{-2t_2^2}{L^2-M^2}).
\end{equation}
\end{widetext}
Now observe that the following is true
\begin{align}
    {L \choose \frac{1}{2}(L+M)} & \exp(\frac{-2t_1^2}{L^{2}+M^{2}-2L}) \exp(\frac{-2t_2^2}{L^2-M^2}) \notag \\ &\leq {L \choose \frac{L}{2}}\exp \left(-\frac{t^{2}_{1} + t^{2}_{2}}{L^{2}} \right),
\end{align}
where $L \in \mathbb{N},  \vert M \vert  \leq L$ and $ t_{1}, t_{2} \in \mathbb{R}^{+}$.

This leads us to the following 
\begin{align}
    \bigcup_{M}\bigcup_{\vert \sigma_{M} \rangle}P\left( \left \vert \sum_{\substack{i,j = 1 \\ i< j}}^{L}\mu_{ij}A^{S}_{ij} -\sum_{\substack{i,j = 1 \\ i< j}}^{L}\mu_{ij}A^{W}_{ij} \right \vert \geq t_{1} + t_{2} \right) \notag \\ \leq 4(L+1) {L \choose \frac{L}{2}}\exp \left(-\frac{t^{2}_{1} + t^{2}_{2}}{L^{2}} \right),
\end{align}
where the union bound has again been used. Now observe that if $t_{1} + t_{2} = \mathcal{O}(L^{\gamma})$ then $t_{1}^{2} + t_{2}^{2} = \mathcal{O}(L^{2\gamma})$, where $\gamma \in \mathbb{R}$. Using the fact ${L \choose \frac{L}{2}} \sim \sqrt{\frac{2}{L\pi}}2^{L}$ for large $L$, and also that using $2^{L} = e^{L{\rm ln}(2)}$, we then arrive at
\begin{align}
    \bigcup_{M}\bigcup_{\vert \sigma_{M} \rangle}P\left( \left \vert \sum_{\substack{i,j = 1 \\ i< j}}^{L}\mu_{ij}A^{S}_{ij} -\sum_{\substack{i,j = 1 \\ i< j}}^{L}\mu_{ij}A^{W}_{ij} \right \vert \geq \mathcal{O}(L^{\gamma}) \right) \notag \\ \leq 2(L+1) \sqrt{\frac{2}{L\pi}}\exp \left(L{\rm ln}(2)-\mathcal{O}(L^{2\gamma - 2}) \right),
\end{align}
which vanishes unless $\gamma \leq \frac{3}{2}$. This leads us to 
\begin{equation}
   \bra{\sigma_1^{\alpha}, ..., \sigma_L^{\alpha}}  D_L^{\alpha}  \ket{\sigma_1^{\alpha}, ..., \sigma_L^{\alpha}} = \mathcal{O}(L^{\frac{1}{2}}) \qquad \forall M, \vert \sigma \rangle.
\end{equation}
Now from Weyl's inequality we know that the eigenvalues of the operator $D_L = H(G_{L}^{S}) - H(G_{L}^{W})$ are therefore bounded as $\mathcal{O}(L^{1/2})$. From here we can invoke Lemma 1 with $\kappa = 1/2$ and Theorem 2 is proven.

\subsection{Appendix B: Analytical solution of the classical Ising model for $W(x,y) = xy$}
We wish to solve Eq. (\ref{Eq:ClassicalIsingEqn}) with $W(x,y) = xy$, i.e. identify the function $\lambda^{z}(x)$ which solves 
\begin{equation}
    \lambda^{z}(x) = \int_{0}^{1}xy\tanh(\beta \lambda^{z}(y) ) dy.
    \label{Eq:SMClassicalIsingEqnXY}
\end{equation}
We start by observing that $\lambda^{z}(x) = f(\beta)x$, where $f(\beta)$ is a real valued function of $\beta$ that is independent of $x$. Substituting this into Eq. (\ref{Eq:SMClassicalIsingEqnXY}) and defining $c = \beta f(\beta)$ gives us
\begin{equation}
    \frac{c}{\beta} = \int_{0}^{1}y\tanh(c y)dy,
    \label{Eq:SMClassicalIsingEqnXYV2}
\end{equation}
We can perform the integration here analytically. First, perform integration by parts and expand into exponentials:
\begin{align}
    &\int_{0}^{1}y\tanh(cy)dy \notag \\ &= \frac{1}{c^{2}} \Bigg(\Biggl[ \frac{1}{2} y^{2} \tanh(cy) \Biggr]_{0}^{c} - \frac{1}{2}\int_{0}^{c}y^{2}\sech^{2}(y)dy \Bigg) \notag \\ &= \frac{1}{c^{2}}\Bigg(\frac{c^{2}\tanh(c^{2})}{2} - \int_{0}^{c}\frac{2y^{2}e^{-2y}}{(1+e^{-2y})^{2}}dy \Bigg).
\end{align}
Now the integral on the RHS can be dealt with by observing that $\frac{e^{-2y}}{(1+e^{-2y})^{2}} = \sum_{n=1}^{\infty}(-1)^{n-1}ne^{-2nx}$, giving us
\begin{align}
    &c^{2}\int_{0}^{1}y\tanh(cy)dy \notag \\ &= \frac{c^{2}\tanh(c^{2})}{2} - \sum_{n=1}^{\infty}(-1)^{n-1}n\int_{0}^{c}2y^{2}e^{-2ny}dy \notag \\ &= \frac{c^{2}\tanh(c^{2})}{2} - \sum_{n=1}^{\infty}\frac{(-1)^{n}(1+e^{-2cn}(1+2cn(1+cn))}{2n^{2}}.
\end{align}

We can evaluate the series by splitting up the numerator and using known results,
\begin{align}
    &c^{2}\int_{0}^{1}y\tanh(cy)dy \notag \\ = \frac{1}{2} - &\frac{\pi^{2}}{24c^{2}} + \frac{1}{c}\ln(1+e^{-2c}) - \frac{1}{2c^{2}}{\rm PL}_{2}(-e^{-2c}).
\end{align}
We can then use this result to reduce Eq. (\ref{Eq:SMClassicalIsingEqnXY}) to Eq. (\ref{Eq:GraphonSolutionIsingXY}) from the main text:
\begin{equation}
    \frac{1}{\beta} = \frac{1}{24c^{3}}\left(12c^{2}-\pi^{2}+24c {\rm ln}(1+e^{-2c}) - 12 {\rm PL}_{2}(-e^{-2c})\right).
    \label{Eq:SMGraphonSolutionIsingXYSM}
\end{equation}
Additionally, it is straightforward to observe that $M^{z} = \int_{0}^{1}\sigma^{z}(x)dx = \int_{0}^{1}\tanh(c x)dx = \frac{\ln(\cosh(c))}{c}$.
\newline 
\subsection{Analytical solution of the ground-state of the transverse field Ising model for $W(x,y) = \sqrt{xy}$}
We wish to solve Eq. (9) in the main text with $W(x,y) = \sqrt{xy}$ and $\beta = \infty$, i.e:
\begin{equation}
    \lambda^{z}(x) = \sqrt{x}\int_{0}^{1}\frac{\sqrt{y}\lambda^{z}(y)}{\sqrt{h^{2}+(\lambda^{z}(y))^{2}}} dy,
    \label{Eq:SMIsingEqn}
\end{equation}
again observing that this implies $\lambda^{z}(x) = \sqrt{x}g(h)$ where $g(h)$ is some real-valued function of $h$ gives 
\begin{equation}
    \int_{0}^{1}\frac{y}{\sqrt{h^{2}+g^{2}(h)y}} dy = 1,
\end{equation}
which we can solve for $g(h)$ (we restrict $h$ and $g(h)$ to be positive real without loss of generality) by a series of substitutions. Yielding
\begin{equation}
    \frac{2(2h^{3} - 2h^{2}\sqrt{h^{2}+g^{2}(h)} + g^{2}(h)\sqrt{h^{2}+g^{2}(h)})}{3g^{4}(h)} = 1,
\end{equation}
which has the solution 
\begin{equation}
    g(h) = \begin{cases}
  \frac{\sqrt{2}}{3}\sqrt{1+(1-3h)\sqrt{1+6h}}  & h < \frac{1}{2}, \\
  0 & \text{otherwise}.
  \label{Eq:SMRootXYg}
\end{cases}
\end{equation}
as in the main text. The transverse and longitudinal magnetisations are determined by the integrals
$M^{x} =  \int_{0}^{1}\frac{h}{\sqrt{h^{2} + g^{2}(h)x}}dx$ and $M^{z} =  - \int_{0}^{1}\frac{g(h)\sqrt{x}}{\sqrt{h^{2} + g^{2}(h)x}}dx$ respectively. The first, by direct integration, yields Eq. (11) in the main text. The second can be done by an extensive series of trigonometric substitutions and results in the following closed form expression
\begin{widetext}
\begin{equation}
    M^{z}(x) = \begin{cases}
        \frac{(1+s)\left(2\sqrt{(1-s)(4+18h^{2}-4s)} - 9h^{2}\left(2\ln(3) + 2\ln(h) - 2\ln(-\sqrt{2-2s} + \sqrt{2 + 9h^{2}-2s})\right)\right)}{108h^{2}(-1 + 2h)} \\
  \qquad \qquad \qquad \qquad \qquad \qquad \qquad \quad  0 & \text{otherwise}
    \end{cases},
\end{equation}
\end{widetext}
where $s = (-1 + 3h)\sqrt{1+6h}$.

\subsection{Appendix C: Further example graphons}
In this section we consider further graphons which were not treated in the main text but frequently appear in the literature on graphons.
\par \textit{Stochastic block models} -
The Stochastic Block Graphon is typically utilised in statistical analysis of networks because they are useful in uncovering clustering in networks \cite{StochasticBlockModel}. The graphon can be expressed as
\begin{equation}
W(x, y) = 
\begin{cases}
p_{11} \quad {\rm if} \quad (x,y) \in X_{1} \times X_1,\\
p_{12} \quad {\rm if} \quad (x,y) \in X_1 \times X_2,\\
...\\
p_{kk} \quad {\rm if} \quad (x,y) \in X_k \times X_k,
\end{cases}
\end{equation}
\noindent with $p_{ij} = p_{ji}$ and the $X_{i}$ specifying disjoint sub-domains of $[0,1]$ such that $\cup_{i=1}^{k} X_{i} = [0,1]$. We write $\Delta X_i$
to indicate the width of the interval $X_i$. 
The continuous mean-field equations then take on the following form

\begin{widetext}
\begin{equation}
    \lambda^{\alpha}(x) = - J^{\alpha}\sum_{j=1}^{k}\int_{X_{j}}\frac{p_{ij}\lambda^{\alpha}(y)\tanh(\beta \sqrt{(\lambda^{x}(y))^{2}+(\lambda^{y}(y)^{2}+(\lambda^{z}(y))^{2}})}{\sqrt{(\lambda^{x}(y))^{2}+(\lambda^{y}(y)^{2}+(\lambda^{z}(y))^{2}}} dy + h^{\alpha} \qquad \forall x \in X_{i}.
    \label{Eq:SMContinuousMFEquationsStochasticGraphon}
\end{equation}
\end{widetext}

\noindent Observe that we can immediately infer from this that $\lambda^{\alpha}(x)$ is constant across each of the domains $X_{i}$. We can thus define $\lambda_i^\alpha = \lambda^{\alpha}(x) \ \forall x \in X_{i}$ and reduce Eq. (\ref{Eq:SMContinuousMFEquationsStochasticGraphon}) to

\begin{align}
    &\lambda^{\alpha}_i \notag \\ &= - J^{\alpha} \sum_{j=1}^{k}\Delta X_j  \frac{p_{ij} \lambda^{\alpha}_j\tanh(\beta \sqrt{(\lambda^{x}_j)^{2}+(\lambda^{y})_j^{2}+(\lambda^{z}_j)^{2}})}{\sqrt{(\lambda^{x}_j)^{2}+(\lambda^{y}_j)^{2}+(\lambda^{z}_j)^{2}}}  + h^{\alpha},
    \label{Eq:SMContinuousMFEquationsStochasticGraphon}
\end{align}
a series of equations which become increasingly complicated to solve as the number of clusters does. In the case of a single cluster we recover the case of an Erdős-R\'enyi graph.

 \par \textit{Growing uniform attachment} -- The growing uniform attachment graphon is given by $W(x, y) = 1 - \max(x, y)$ \cite{PerferentialAttachmentGraphon}. The graphs which are finite realisations of this graphon will consist of nodes in which the average connectivity of a node varies uniformly across the graph. Such graphs are therefore highly inhomogenous in their average vertex connectivity.
Substituting $W(x, y) = 1 - \max(x, y)$ into Eq. (\ref{Eq:ContinuousMFEquations}) and differentiating the left and right hand sides twice with respect to $x$ leads us to the following coupled second-order ODEs

\begin{equation}
    \frac{d^2 \lambda^{\alpha}(x)}{dx^2} = J^{\alpha} \frac{\lambda^{\alpha}(x)\tanh{ (\beta \sqrt{\lambda^x(x)^2 + \lambda^y(x)^2 + \lambda^z(x)^2})}}{\sqrt{\lambda^x(x)^2 + \lambda^y(x)^2+ \lambda^z(x)^2}},
\end{equation}
with $\alpha = x,y,z$, boundary conditions $\lambda^{\alpha}(1) = 0$ and  $\frac{d\lambda^{\alpha}(x)}{dx}\vert_{x=0} = 0 \ \forall \alpha$.

\par \textit{Maximally irregular graph} -- The maximally irregular graph is the finite connected graph where each site (other than one pair) has a different co-ordination number to any others \cite{Tindall2022}. Taking the thermodynamic limit of the adjacency matrix results in the graphon
\begin{equation}
    W(x,y) = 
    \begin{cases}
        1 \quad x + y \leq 1 \\
        0 \quad {\rm otherwise.}
    \end{cases}
\end{equation}
and the integral equations in Eq. (\ref{Eq:ContinuousMFEquations}) reduce (upon differentiation) to the following three coupled first order ODEs
\begin{widetext}
\begin{equation}
    \frac{d\lambda^{\alpha}(x)}{dx} = - J^{\alpha}\frac{\lambda^{\alpha}(1-x) {\rm \tanh}(\beta \sqrt{(\lambda^{x}(1-x))^{2}+(\lambda^{y}(1-x))^{2}+(\lambda^{z}(1-x))^{2}}}{\sqrt{(\lambda^{x}(1-x))^{2}+(\lambda^{y}(1-x))^{2}+(\lambda^{z}(1-x))^{2}}}, \qquad \alpha = x,y, z,
\end{equation}
with boundary conditions $\lambda^{\alpha}(1) = 0 \ \forall \alpha$. Such equations are known as functional differential equations and have been studied extensively in both Mathematics and the applied Sciences \cite{FunctionalDifferentialEquations}.
\end{widetext}

\subsection{Appendix D: Numerical details}

\par
\begin{figure}[t!]
    \includegraphics[width =\columnwidth]{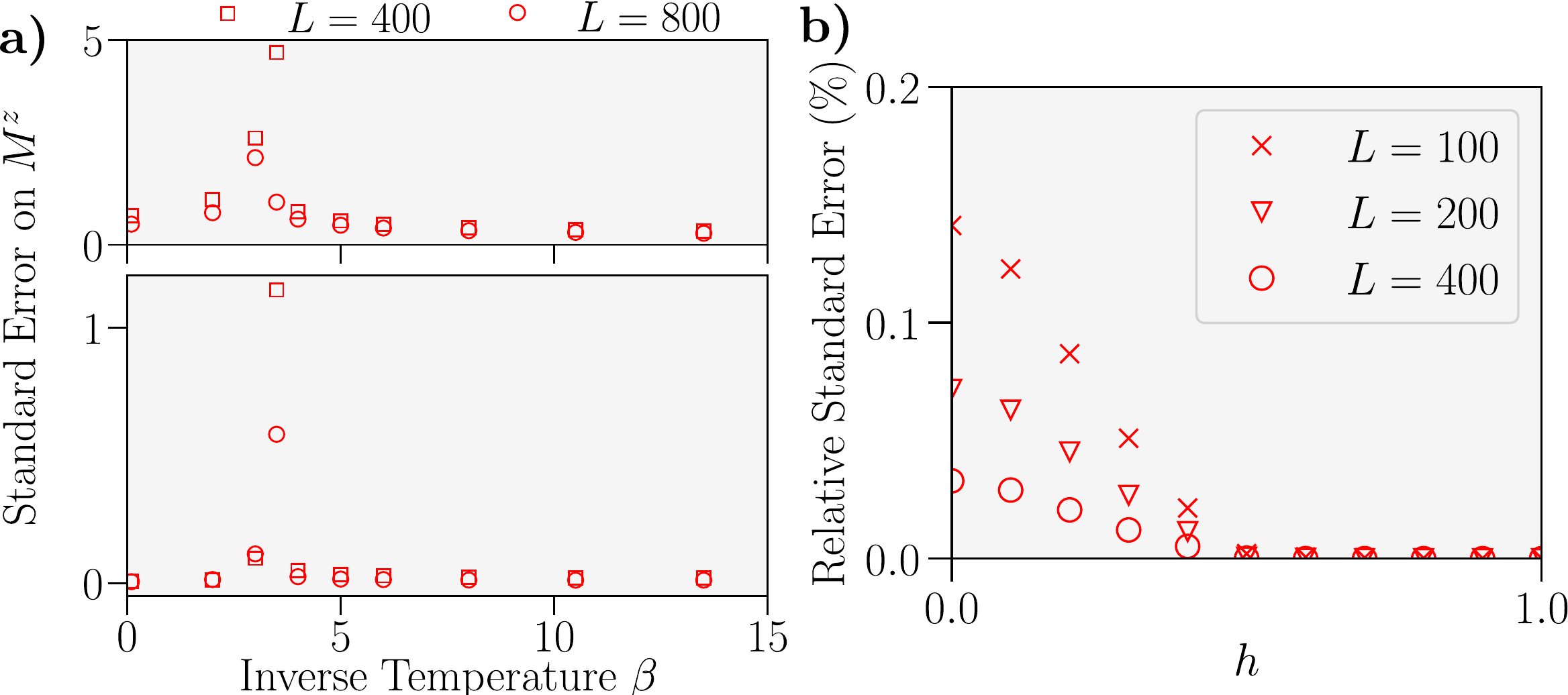}
    \caption{a) Standard error on the mean for the Monte-Carlo calculations of $M^{z}$ for the classical Ising Model on stochastic finite-size realisations of $W(x,y) = xy$. Top plot is standard error on the mean from $5000$ Monte-Carlo samples of $M^{z}$ at a given $\beta$ and $L$. Data points are averaged over $100$ stochastic realisations of $W(x,y)$. Bottom plot is the standard error on the mean from the $100$ stochastic realisations of $W(x,y)$, averaged over the $5000$ samples taken for each realisation. b) Relative standard error on the mean (standard error on the mean as a percentage of the mean) for the ground state energy of the transverse field Ising model calculated via DMRG. Standard error is that originating from the $100$ stochastic realisations of $W(x,y) = \sqrt{xy}$ at a given $h$ and system size $L$.}
    \label{Fig:SF1}
\end{figure}

 \textit{Classical Ising model} --- For the finite-size data plotted in Figure 1 of the main text we used Monte-Carlo simulations. Specifically, for a given $L$ we drew a finite random-exchange realisation of the graphon $W(x,y) = xy$ and for a given temperature $\beta$ utilised the Metropolis-Hastings algorithm to generate $N_{\rm Samples} = 5000$ for the Magnetisation Density $M^{z}$. We used a Markov chain length of $250$ between each sample and threw away the first $1000$ samples. For each $L$ we took $100$ stochastic realisations of the graphon $W(x,y)$ and averaged our results over this. There are thus two sources of statistical error in our simulations: the error from sampling a finite number of stochastic realisations and the error from taking a finite number of Monte-Carlo samples. In Fig. \ref{Fig:SF1} we plot the standard error on the mean from both of these sources, the values are negligible in comparison to the scale ($0 \rightarrow 1$) of Fig. \ref{Fig:F1} in the main text.
\par \textit{Transverse Ising model} --- For the data plotted in Figure 2 of the main text we used the Density-Matrix-Renormalisation-Group (DMRG) algorithm to find the ground-state of the transverse field Ising model. For a given $L$ we drew a finite stochastic realisation of the graphon $W(x,y) = \sqrt{xy}$. Then, for a given field strength $h$ we took a random Matrix Product State with a small bond dimension $\chi$ and successively performed DMRG sweeps, letting the bond-dimension double every $4$th sweep until the energy converges to within $0.1\%$ of that for the previous bond dimension. There is thus only one source of statistical error in this simulation: the error from sampling a finite number ($100$) of stochastic realisations. In Fig. \ref{Fig:SF1} we plot this error as a percentage and observe that it is on the order of $0.1\%$. The ordering of the sites (from left to right) of the Matrix Product State was taken to be identical to the ordering $v = 1...L$ of the sites of the graph.

\subsection{Appendix E: The Graphon as the Limit Object of Dense Graph Sequences}

We provide mathematical details on how the graphon $W$ is the limit object of a sequence of dense graphs $(G_n)_{n \in \mathbb{N}}$ where $n$ is the number of vertices. This Appendix closely follows Ref. \cite{Lovasz_2012}, although the theory on graph limits was first developed in Ref. \cite{Graphon1}. The interested reader should consult either of these for more detail.

Consider two simple graphs $F$ and $G$, where we define the number of vertices of $F$ to be $k$ and that of $G$ to be $n$. A homomorphism from $F$ to $G$ is a map which preserves edges. This means that given an edge $(i, j) \in E(F)$ (here $E(F)$ is the edge set of $F$), and a homomorphism $h$, there is always an edge $(h(i), h(j)) \in E(G)$--- the set of edges of $G$.
Let ${\rm hom}(F, G)$ indicate the number of homomorphisms from $F$ into $G$. The homomorphism density $t(F, G)$  is then defined to be
\begin{equation}
    t(F, G) = \frac{{\rm hom}(F,G)}{n^k}.
\end{equation}
The homomorphism density is the probability of a random map from the graph $F$ to the graph $G$ being a homomorphism, since $n^k$ is the total number of maps from a graph with $k$ vertices to a graph with $n$ vertices.

Suppose that instead we are given a graphon, such as $W_G$ --- the stepped graphon corresponding to the graph $G$ which is defined as $W_{G}(x,y) = A_{v,v'}$ for $(x,y) \in [(v-1)/n, v/n] \times [(v'-1)/n, v'/n]$ (with $A$ being the adjacency matrix of $G$). In this case, the homomorphism density is defined to be 
\begin{equation}
    t(F, W_G) = \int_{[0,1]^{k}} \prod_{(i, j) \in E(F)} W(x_i, x_j) \prod_{i \in 1:k} dx_i
\end{equation}
Here the same definition holds for any arbitrary graphon $W$.

\par The homomorphism density with reference to a finite graph $F$ indicates the relative likelihood of the graph $G$ or more generally graphon $W$ containing an instance of $F$ inside of it. If two graphs or graphons have similar homomorphism densities for $\textit{all}$ simple graphs $F$, then these graphs are similar. The definition of convergence of a sequence of graphs hinges precisely on this concept.

\begin{definition}[Convergent Graph Sequence]
    A sequence $\left(G_n \right)$ of simple graphs with $V(G_n) \rightarrow \infty$ as $n \rightarrow \infty$ converges if the subgraph densities $t(F, G_n)$ converge for all simple graphs $F$.
\end{definition}

The above definition gives allows us to precisely define in what sense $W$ can be considered a limit object.

\begin{theorem}[Lovasz, 2012 \cite{Lovasz_2012}]
    Let $(G_n)$ be a sequence of simple graphs with $V(G_n) \rightarrow \infty$. If $(G_n)$ converges, there exists a graphon $W$ such that $t(F, G_n) \rightarrow t(F, W)$ for all simple graphs $F$.
\end{theorem}

The above theorem tells us that if the sequence $(G_n)$ converges, then there exists some limit object---the graphon---which captures the limiting homomorphism density counts of the sequence of graphs for \textit{all} simple graphs.

\par There is a second, equivalent, definition of convergence which us allows us to define $W$ as an appropriate limit of a sequence of dense graphs. This definition utilises the cut distance of two graphs. 

\begin{definition}[Cut Distance]
Given two graphons $W$ and $W^{\prime}$, define the cut distance between them to be

\begin{align}
    \delta_{\square}(&W, W^{\prime}) \notag \\ &: = \inf_{\phi, \psi} \sup_{S, T} \left|\int_{S \times T} (W(\phi(x), \phi(y)) - W^{\prime}(\psi(x),\psi(y)))\right|
\end{align}
where the infimum is taken over all vertex re-labelings $\phi$ of $W$ and $\psi$ of $W^{\prime}$. The supremum is taken over all measurable subsets $S$ and $T$ of $\left[0, 1\right]$. 
\end{definition}

The cut distance is a metric on the space of graphons (up to weak isomorphism). It maximises the difference between the integral of the two graphons on measurable intervals $S$ and $T$ which together form a box $S \times T$. This step can be thought of as maximising the difference in edges between those vertices contained in $S \times T$. The infimum is then taken on that chosen interval over all measure preserving maps, in order to ensure that the cut distance is zero for weakly isomorphic graphons. The following theorem can then be proven from the above definitions.

\begin{theorem}[Lovasz, 2012 \cite{Lovasz_2012}]
    Given a sequence $\left(G_n \right)$ of simple graphs with $|V(G_n)| \rightarrow \infty$ as $n \rightarrow \infty$, the sequence is said to converge to the graphon $W$ if $\delta_{\square}(W_{G_n}, W) \rightarrow 0$ as $n \rightarrow \infty$.
\end{theorem}

This theorem provides alternative definition for the graphon as a limit object. In this definition, we envisage instead the pixelated adjacency matrix of the sequence of simple graphs $G_n$ approaching (via the cut distance) that of the limit object $W$.

Importantly, the above definitions and theorems can be generalised to sequences of weighted graphs by requiring the graphs to have uniformly bounded edgeweights. Moreover, we emphasise that these limits only make sense for sequences of \textit{dense} graphs, because it can be shown that sparse graph sequences always have as their limit the graphon $W(x,y) = 0$ for all $x$ and $y$.

\end{document}